\documentclass{aa}
\usepackage{graphicx}
\usepackage{txfonts}
\usepackage{epsfig}

\begin{document}

\title{Red giant branch stars as probes of stellar 
populations. II. Properties of the newly discovered 
globular cluster GLIMPSE-C01}
\subtitle{}

\author{Valentin D. Ivanov\inst{1}
  \and Radostin Kurtev\inst{2}
  \and Jordanka Borissova\inst{1}
}

\offprints{V.\,D.~Ivanov}

\institute{European Southern Observatory, Ave. Alonso de 
           Cordova 3107, Casilla 19, Santiago 19001, Chile,
           vivanov@eso.org, jborisso@eso.org
  \and Departamento de F\'isica y Meteorolog\'ia, Facultad 
       de Ciencias, Universidad de Valpara\'{\i}so, 
       Ave. Gran Breta\~na 644, Playa Ancha, Casilla 53,
       Valpara\'iso, Chile, rkurtev@uv.cl}

\date{Received ... / Accepted ...}

\titlerunning{Near-IR photometry of GLIMPSE-C01}
\authorrunning{Ivanov et al.}

\abstract{
Deep near infrared photometry of the newly discovered 
Galactic globular Cluster GLIMPSE-C01 is reported. We 
derived for the first time the metal abundance of this 
object from the slope of the RGB: [Fe/H]=$-$1.61$\pm$0.14 
in the scale of Zinn (as implemented in Harris \cite{har96}), 
[Fe/H]=$-$1.44$\pm$0.12 in the scale of Caretta \& Gratton 
(\cite{car97}), and [Fe/H]=$-$1.12$\pm$0.12 in the scale of 
Ferraro et al. (\cite{fer99}). The tip and the clump of the 
red giant branch were used to confirm the estimates of 
Kobulnicky et al. (\cite{kob05}), placing the cluster at 
D$\sim$3.7$\pm$0.8 kpc, behind A$_V$$\sim$15 mag of visual 
extinction. The best fit to the radial surface brightness 
profile with a single-mass King's model (\cite{kin62}) 
yielded core radius r$_c$=0.78 arcmin, tidal radius r$_t$=27 
arcmin, and central concentration c=1.54.

Finally, we estimate the number of the ``missing'' globulars 
in the central region of the Milky Way. Based on the spatial 
distribution of the known clusters, and assuming radial 
symmetry around the Galactic center, we conclude that the 
Milky Way contains at least 10$\pm$3 undiscovered objects. 
The distribution of known clusters in the bulge seem to 
resemble the orientation of the Milky Way bar.

\keywords{Galaxy: globular clusters: general - Galaxy: 
          abundances - galaxies: abundances - galaxies: 
          distances and redshifts - stars: distances - stars: 
          abundances}
}

\maketitle

\section{Introduction}
The all-sky infrared surveys carried out during the recent years 
have brought the discovery of a number of new clusters, hidden by
the dust extinction in the plane of the Milky Way. They usually 
suffer A$_V$$\geq$10-20 mag of extinction, making them invisible 
in the optical wavebands. The vast majority of these objects 
appear to be a few million years old (Ivanov et al. \cite{iva02}, 
\cite{iva05}, Borissova et al. \cite{bor03}, \cite{bor05}) but a 
few have proved to be analogues of ``classical'' globular 
clusters (Hurt et al. \cite{hur00}, Kobulnicky et al. \cite{kob05}, 
Carraro \cite{car05}).

The globular cluster GLIMPSE-C01 was discovered (Kobulnicky et 
al. \cite{kob05}) during the Galactic Legacy Infrared Mid-Plane 
Survey Extraordinaire (hereafter GLIMPSE; Benjamin et al. 
\cite{ben03})). The survey is mapping the Galactic plane from 
$\mid$l$\mid$ = 10$\degr$ to 65$\degr$ and 
$\mid$b$\mid$$\leq$1$\degr$ with the Infrared-Array Camera (IRAC; 
Fazio et al. \cite{faz04}) on {\it Spitzer Space Telescope} in 
3.6, 4.5, 5.8, and 8.0 micron bands. It was also identified 
independently by Simpson \& Cotera (\cite{sim04}) who 
cross-correlated {\it ASCA} X-ray and {\it IRAS} sources with 
the 2\,MASS.

The isochrone comparison indicated that GLIMPSE-C01 is indeed a 
globular cluster, at least a few Gyr old. The near infrared 
color-magnitude diagram (CMD) suggested that the object suffers 
A$_V$=15$\pm$3 mag of extinction, and the analysis of the
line-of-sight $^{13}$CO yielded a distance of 3.1-5.2 Kpc, based 
on a cinematic model of the Milky Way. The question of the 
cluster abundance remained open, as well as the need to verify
the distance.

Here we report properties -- metallicity, extinction, and 
distance -- derived from deep near infrared photometry of 
GLIMPSE-C01. We also estimate the possible number of undiscovered 
globular clusters in the central region of the Milky Way.

\section{Observations and data reduction}

The $JHK_S$ imaging observations of GLIMPSE-C01 were carried 
out in Nov 2004 under non-photometric condition and 1 arcsec 
seeing with the SofI (Son of ISAAC) at the NTT. The instrument is 
equipped with Hawaii HgCdTe 1024$\times$1024 detector, with pixel 
scale of 0.288 arcsec px$^{-1}$. 

We took 12 images in each filter, in jittering mode with 3 arcmin 
jitter box size to ensure that on each image the cluster is 
located on a different place on the array. Each image was the 
average of 3 frames of 10 sec for $J$, 13 frames of 3 sec for $H$, 
and 20 frames of 3 sec for $K_S$, comprising total integration 
times of 6, 7.8, and 12 min, respectively. Individual images were 
sky-subtracted, flat-fielded, aligned, and combined into a single 
image. 

The stellar photometry of the final images was carried out with
{\sc ALLSTAR} in {\sc DAOPHOT II} (Stetson \cite{ste93}). The 
typical photometric errors vary from 0.03 mag for the 
K$_S$$\sim$10 mag stars to 0.10 mag for K$_S$$\sim$13 and 0.15 
mag for K$_S$$\sim$16 mag. The photometry of the the faint stars 
is affected by confusion, especially near the cluster center but 
our metallicity estimates is based on stars with K$_S$$\leq$13 
mag.

The photometric calibration was performed by comparing our 
instrumental magnitudes with the 2\,MASS measurements of 67 
stars, covering the color ranges 0.61$\leq$$J$$-$$K_S$$\leq$5.98 
and 0.21$\leq$$H$$-$$K_S$$\leq$2.32 mag, and magnitude range 
8.33$\leq$$K_S$$\leq$13.13 mag. The transformation 
equations are:
\begin{equation}
J-K_S=(0.981\pm0.005)\times(j-k)+(3.208\pm0.007), 
\end{equation}
\begin{equation}
H-K_S=(0.971\pm0.016)\times(h-k)+(1.820\pm0.014),
\end{equation}
and
\begin{equation}
K_S=(1.008\pm0.007)\times\,k+(0.0126\pm0.006)\times(J-K_S)-
(10.096\pm0.131),
\end{equation}
with r.m.s.=0.047, 0.062, and 0.044 mag, respectively. Here $k$, 
$j$$-$$k$, $h$$-$$k$ are the instrumental magnitudes and colors 
and $K_S$, $J$$-$$K_S$, $H$$-$$K_S$ are the magnitudes and 
colors in the 2\,MASS system. The standard error values are 
given after each coefficient. 

A true-color composite of GLIMPSE-C01 is shown in 
Figure~\ref{GLIMPSE-C01_RGB}. The image indicates substantial 
variations of the extinction in the field. Dust structures are 
present even across the cluster face, suggesting that the 
completeness of the photometry varies significantly. The 
crowding also contributes to this effect, as it can be seen 
from the luminosity function of stars in the field of 
GLIMPSE-C01 (Figure~\ref{GLIMPSE-C01_LF}): it is clear that the 
completeness of the photometry in the cluster area (long-dashed 
line) is at least a magnitude shallowed than that in the field 
(dotted line).

\begin{figure}
\caption{True-color near infrared composite image of GLIMPSE-C01. 
The $J$, $H$, and $K_S$ bands correspond to blue, green and red, 
respectively. North is up, and East is to the left. The field of 
view is 4.92 arcmin on the side. The dark lane in the North-West 
part of the image corresponds to a region of mid-infrared 
emission, and it is a real feature. (This figure is available in 
color from the electronic edition of the journal.)}
\label{GLIMPSE-C01_RGB}
\end{figure}

\begin{figure}
\resizebox{\hsize}{!}{\includegraphics{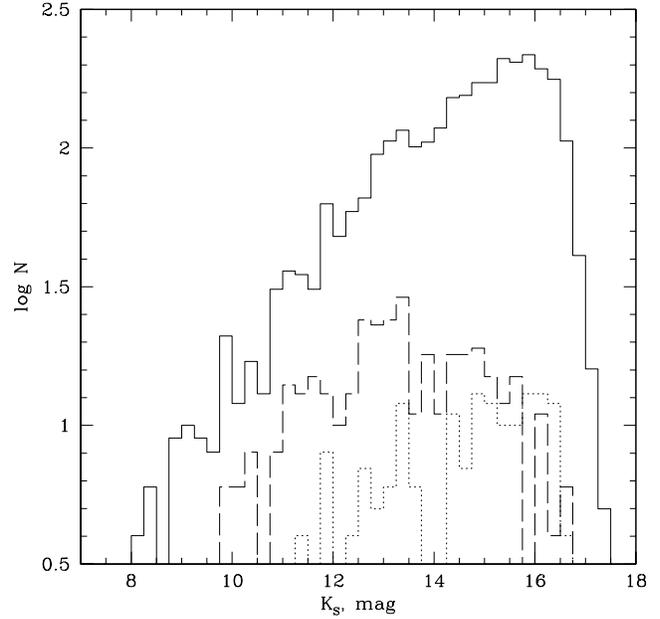}}
\caption{Near infrared luminosity function for: all stars in the 
field of GLIMPSE-C01 (solid line), stars within R$\leq$30 arcsec 
from the center of the cluster (long-dashed line), and stars in 
an annulus with inner radius of 50 arcsec and the same area as the
cluster region (dotted line).}
\label{GLIMPSE-C01_LF}
\end{figure}

\section{Properties of GLIMPSE-C01}

\subsection{Structural parameters}

\begin{figure}
\resizebox{\hsize}{!}{\includegraphics{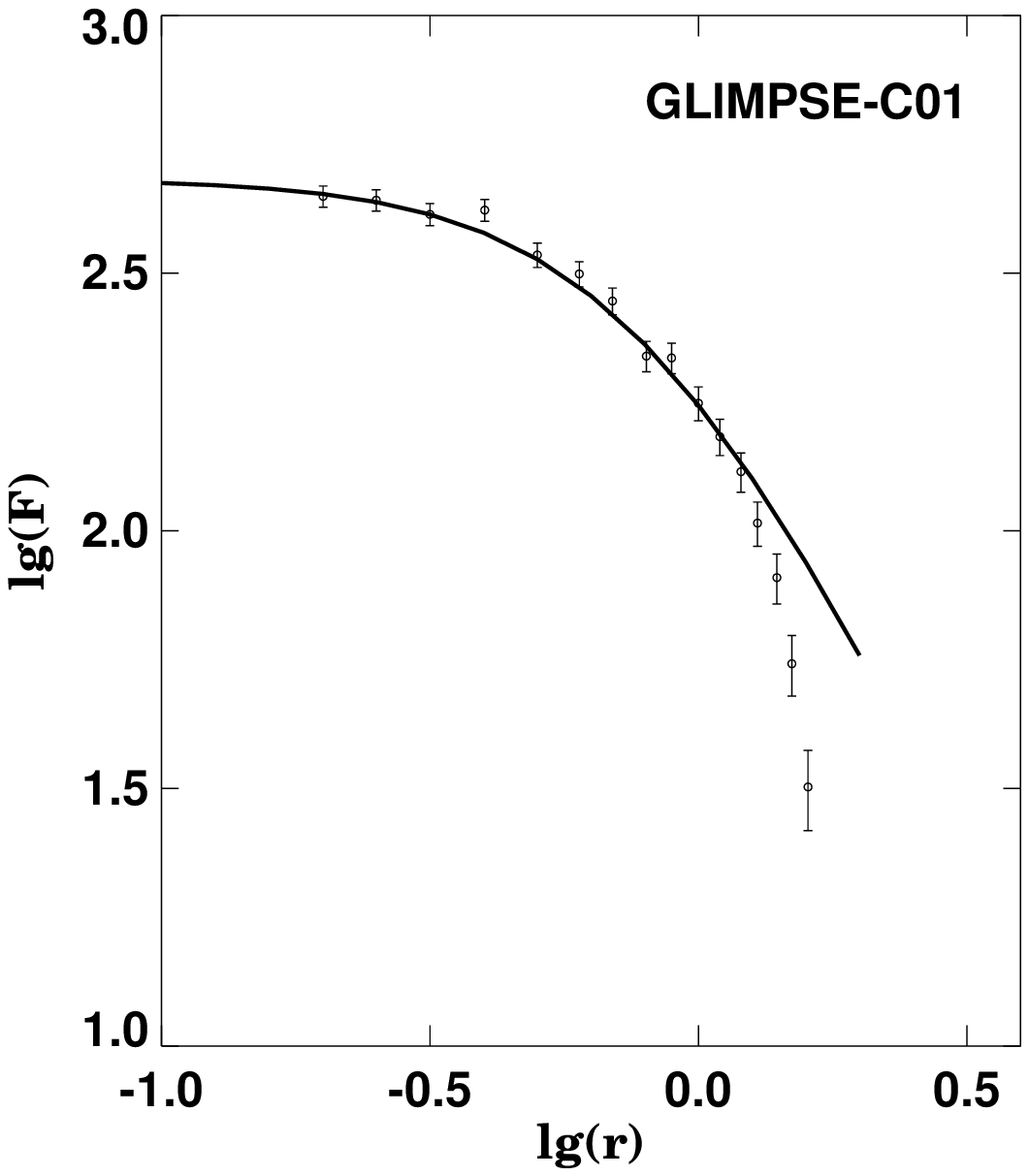}}
\caption{Structural parameters of the cluster: King's profile 
fit.}
\label{KingsProfile}
\end{figure}

The structural parameters of GLIMPSE-C01 were determined using 
the iterative star counts method of King (\cite{kin62}) after 
randomly removing the field stars (see Sec.~\ref{Section_MC}).
The radial profile was built only from the stars in the 
South-Eastern half of the GLIMPSE-C01 to exclude the effects 
of the variable extinction across the cluster face. 
The best fit to the radial surface brightness profile with a 
single-mass King's model (Figure~\ref{KingsProfile}) yielded 
core radius r$_c$=0.78 arcmin, tidal radius r$_t$=27 arcmin, 
and central concentration c=1.54. The inhomogeneous foreground 
extinction -- significant even in the near-infrared -- makes 
it impossible to confirm the suggestion of Kobulnicky et al. 
(\cite{kob05}) that GLIMPSE-C01 is elongated but we certainly 
can not exclude such a possibility.

\subsection{Distance and extinction}

\begin{figure*}
\resizebox{\hsize}{!}{\includegraphics{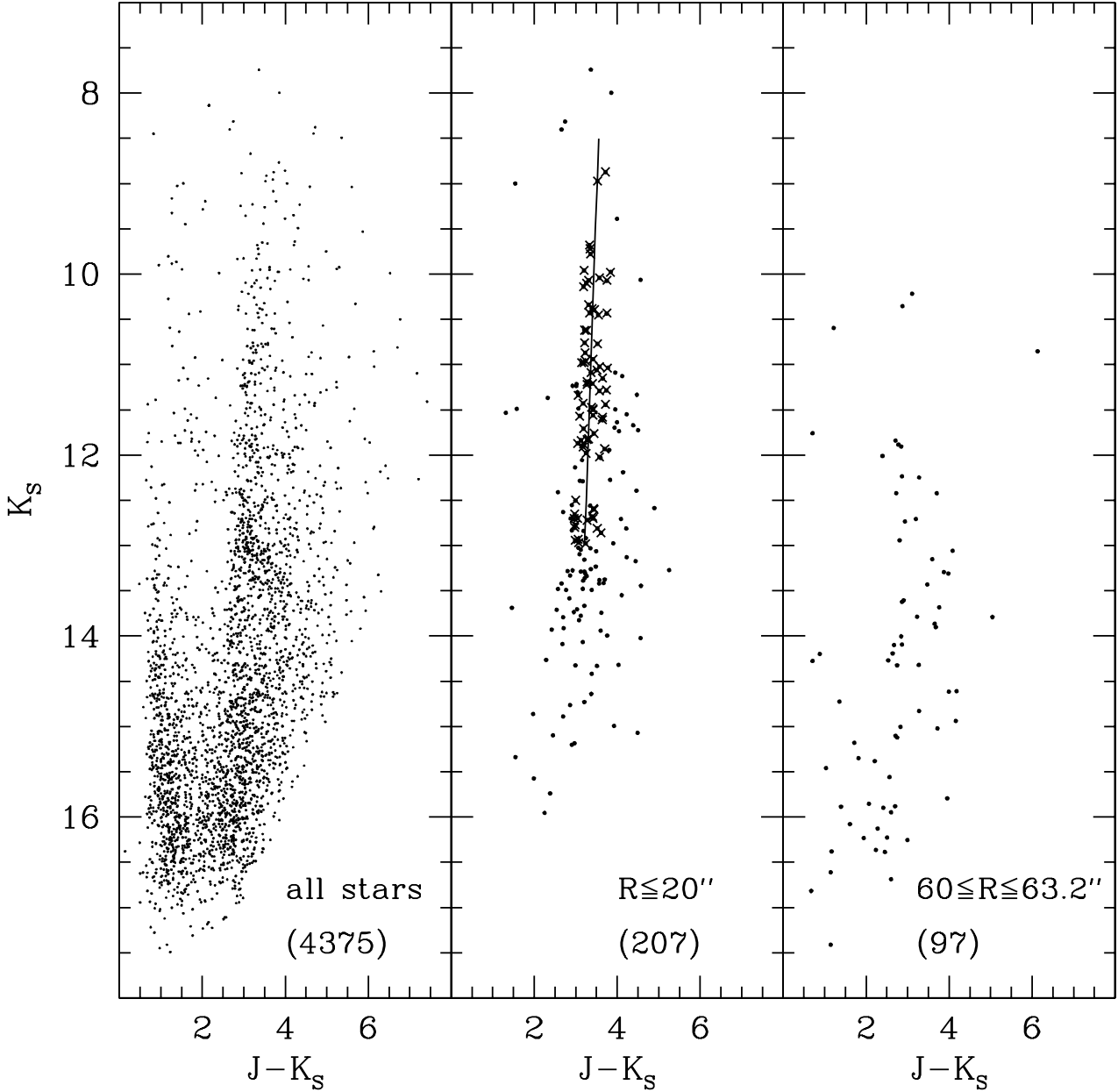}}
\caption{Near infrared color-magnitude diagram for: all stars in 
the field of GLIMPSE-C01 (left panel), and for starts in two 
regions with identical areas -- a circle with radius R=20 arcsec 
around the cluster center (middle panel), and an annulus around 
the cluster with inner radius of 60 arcsec (right panel). The 
numbers of stars plotted in each panel are given in brackets. In 
the middle panel: the best fit to the RGB is drawn with a solid 
line; crosses mark stars included in one realization (see 
Section~\ref{Section_MC} for explanation).}
\label{GLIMPSE-C01_CMD}
\end{figure*}

The CMD of GLIMPSE-C01 is plotted in 
Figure~\ref{GLIMPSE-C01_CMD}. The left panel contains all stars 
with $J$ and $K_S$ photometry, and shows the presence of a 
few sequences. The nature of these sequences become apparent 
from the other two panels: the cluster (middle panel) shows a 
red giant branch (hereafter RGB) at $J$$-$$K_S$$\sim$3-4 mag, 
and the field (right panel) is dominated by main sequence at 
$J$$-$$K_S$$\sim$1-2 mag and a group of red stars with 
$J$$-$$K_S$$\geq$3 mag, possible just highly reddened 
background objects.


An inspection of the CMD reveals the presence of a clump of 
stars in the red giant at $K_S$$\sim$13.0 mag. It is also 
noticeable on the cluster luminosity function 
(Figure~\ref{GLIMPSE-C01_LF}). Taking into account the width 
of the structure along the $K_S$ axis, we assign an uncertainty 
of 0.25 mag (equal to the bin size used to build the luminosity 
function) to the apparent magnitude of the clump. If this 
structure is indeed the red clump, it has an absolute magnitude 
of $M_K$=$-$1.61$\pm$0.03 mag (Alves \cite{alv00}), yielding a 
distance modulus $(m$$-$$M)_K$$\sim$14.6$\pm$0.3 mag. Note that 
this calibration doesn't take into account any metallicity 
effects (see for discussion Salaris \& Girardi \cite{sal02}).

The distance to the cluster can also be measured using the RGB 
tip (i.e. Ivanov \& Borissova \cite{iva02}) but this method 
suffers from strong metallicity dependence and it is hampered 
by the small number of stars at the upper end of the RGB. 
Nevertheless, we carried out this test as a consistency check.
If the tip is located at $K_S$$\sim$8.7 mag, and assuming 
$M_K$=$-$6.1 mag for [Fe/H]=$-$1.5 (Ivanov \& Borissova 
\cite{iva02}) we obtain $(m$$-$$M)_K$$\sim$14.8 mag, in 
agreement with the estimate given above.

\begin{figure}
\resizebox{\hsize}{!}{\includegraphics{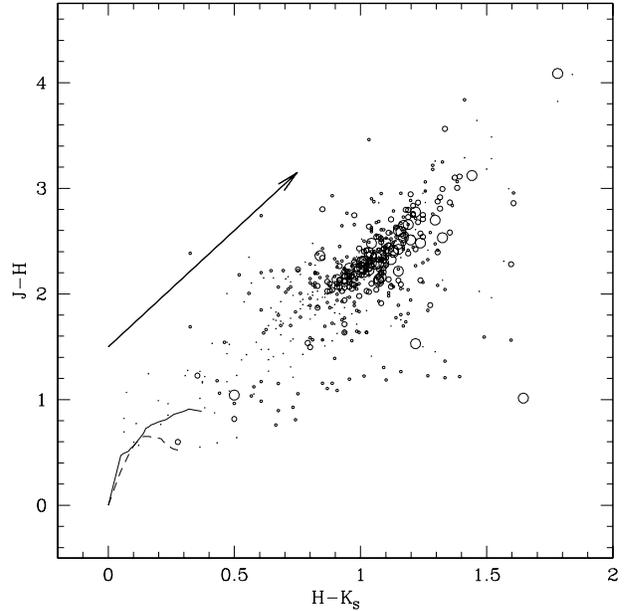}}
\caption{Near infrared color-color diagram for starts in two 
regions in the vicinity of GLIMPSE-C01, with identical areas -- 
a circle with radius R=30 arcsec around the cluster center 
(open circles; the size is proportional to the apparent 
brightness of the star), and an annulus around the cluster with 
inner radius if 50 arcsec (solid dots). A vector representing 
visual extinction A$_V$=15 mag is plotted. The colors of RGB 
(solid line) and the main sequence stars (dashed line) are 
also shown (Frogel et al. \cite{fro78}; no correction for the 
photometric systems is applied).}
\label{GLIMPSE-C01_CCD}
\end{figure}

The color-color diagram of the cluster field is shown in 
Figure~\ref{GLIMPSE-C01_CCD}. Our data confirm the visual 
extinction A$_V$$\sim$15$\pm$3 mag derived by Kobulnicky et al. 
(\cite{kob05}). Throughout this paper we used the reddening law
of Rieke and Lebofsky (\cite{rie85}), giving 
A$_K$$\sim$1.7$\pm$0.3 mag. This yields reddening-free distance 
modulus of $(m$$-$$M)_0$$\sim$12.9$\pm$0.4 mag, or 
D$\sim$3.8$\pm$0.7 kpc, in agreement with Kobulnicky et al. 
(\cite{kob05}). The CMD shows a notable spread of colors among 
the fainter stars that might be contributed to differential 
extinction and to contamination from extended emission and  
background sources.

\subsection{Metal abundance\label{Section_MC}}

The RGB slope allows to derive abundances of globular clusters 
(i.e. De Costa \& Armandroff \cite{dac90}) because of the 
metallicity-dependent opacities in cool stars. The RGB slope 
is independent of the reddening, which is an important 
advantage in studies of heavily obscured clusters such as 
GLIMPSE-C01. Here we applied the analysis and the calibrations, 
described in Ivanov \& Borissova (\cite{iva02a}). 

\begin{figure}
\resizebox{\hsize}{!}{\includegraphics{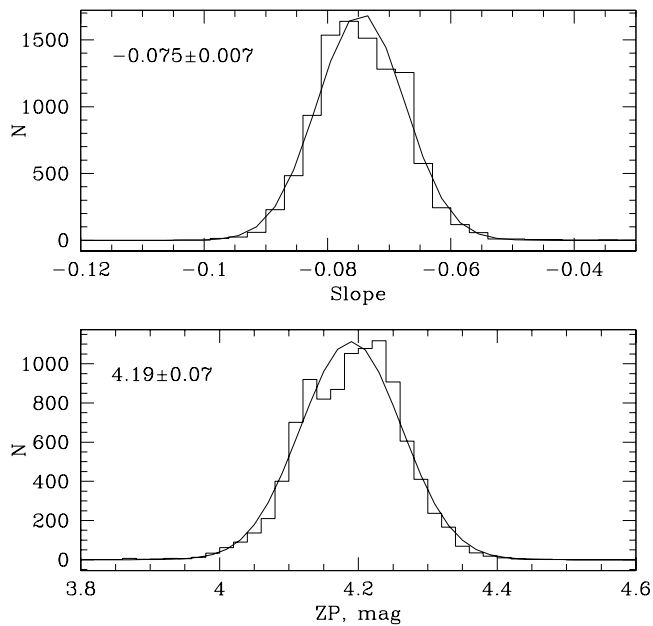}}
\caption{Distributions of the RGB slope (top panel) and the 
zero point (bottom panel) of the fits for the 10$^4$ 
Monte-Carlo realizations with random removal of the fore- 
and background contamination. Gaussian fits to the 
distributions are overplotted, and the average values with 
their standard errors are given. See Section~\ref{Section_MC} 
for details.}
\label{GLIMPSE-C01_RGB_fits}
\end{figure}

The first step was to remove the fore- and background stars, 
which constituted $\sim$8-9\% of the stars in the designated 
cluster area -- a circle with 40 arcsec diameter. We used a 
Monte-Carlo technique: the CMD of the cluster was divided into 
rectangular bins. Similarly, the CMD of a circular annulus, 
centered on the cluster, with an inner radius 60 arcsec, and 
the same are as the cluster region. Then, from each bin of 
the cluster CMD we removed randomly a number of star, equal to
the number of stars in the corresponding bin of the ``field'' 
CMD, producing a single realization. Next, we carried out a 
least square fit to the RGB (Ivanov \& Borissova \cite{iva02a}; 
see also Appendix~\ref{App} in this work):
\begin{equation}
(J-K_S) = ZP + Slope \times K_S
\end{equation}
on the corrected diagram to derive the slope and the zero point. 
The RG locus was 
defined after inspecting the CMD: 8.5$\leq$$K$$_S$$\leq$13.0 mag, 
and 2.5$\leq$$J$$-$$K$$_S$$\leq$4.0 mag. The bin-size was 0.5 
mag along both axis. We removed the 10-$\sigma$ outliers, and 
repeated the fitting. Typically, the fitting coefficients 
obtained in the two iterations were statistically 
indistinguishable. Note that the spread of colors along the 
RGB is dominated by differential extinction, rather than 
measurement errors. Total of 10$^4$ realizations were obtained, 
and we averaged the fitting results. The distributions of the 
RGB slope and the zero point of the fits are shown in 
Figure~\ref{GLIMPSE-C01_RGB_fits}.

Finally, we derived the metal abundance from the calibration 
of Ivanov \& Borissova (\cite{iva02a}), and obtained 
[Fe/H]=$-$1.61$\pm$0.14 in the scale of Zinn (as implemented
in Harris \cite{har96}), 
[Fe/H]=$-$1.44$\pm$0.12 in the scale of Caretta \& Gratton 
(\cite{car97}), and 
[Fe/H]=$-$1.12$\pm$0.12 in the scale of Ferraro et al. 
(\cite{fer99}). The calibrations of Valenti, Ferraro, \& 
Origlia (\cite{val04}) wield [Fe/H]=$-$1.14$\pm$0.16 and 
[Fe/H]=$-$0.97$\pm$0.15, for the last two scales. Both values 
are systematically lower than ours but the differences are 
within 1-2\,$\sigma$. 

These measurements rely on the important assumption that 
GLIMPSE-C01 has the same age as the other Milky Way 
globular clusters, used to derive the metallicity versus the
RGB slope calibrations. The comparison of the cluster CMD 
with stellar isochrones (Kobulnicky et al. \cite{kob05}) 
seems to indicate an age of at least 8 Gyr but this issue 
can not be addressed until more accurate age estimate 
becomes available.

\section{The case of the missing 
clusters\label{Missing_clusters_section}}

The discoveries of GLIMPSE-C01 and Whitting-1 prompted us to 
reinvestigate the question if there are any more undiscovered 
globular clusters in the Milky Way. The total number of Milky 
Way globulars estimated by van den Bergh (\cite{van98}) is 
160$\pm$20, slightly above the currently known $\sim$150 
objects. Barbuy, Bica, \& Ortolani (\cite{bar98}; see their 
Figure 3 and 4) argued that some missing clusters are probably 
located in the general direction of the Galactic Center. Here 
we adopt a modification of their method. The XYZ Galactic 
coordinates used in this section are taken from Feb 2003 
edition of the Milky Way Globular Cluster Catalog (Harris 
\cite{har96}) with their face value. A possible caveat is the
variation of the ratio of the total extinction to the reddening 
R$_V$. For example, a change or R$_V$ from 3 to 3.3 introduces 
10\% change in the distances, or almost 1 kpc at the distance 
of the Galactic Center. The averaging over different dust 
properties along the line of sight helps to minimize the 
related distance uncertainties but the variations of R$_V$ 
toward the bulge definitely need further investigation.

First, we assumed that all missing clusters are located in 
the Galactic plane and close to the Galactic Center. 
Therefore, we consider only the region with 
$\mid$Z$\mid$$\leq$0.5 kpc and R$_{GC}$$\leq$3.0 kpc. These 
constraints reflect the spatial distribution of the obscuring 
material. They make our estimate of the missing clusters only 
a lower limit, as it was demonstrated by the recent discovery 
of the off-the-plane cluster Whiting-1. 

\begin{figure}
\resizebox{\hsize}{!}{\includegraphics{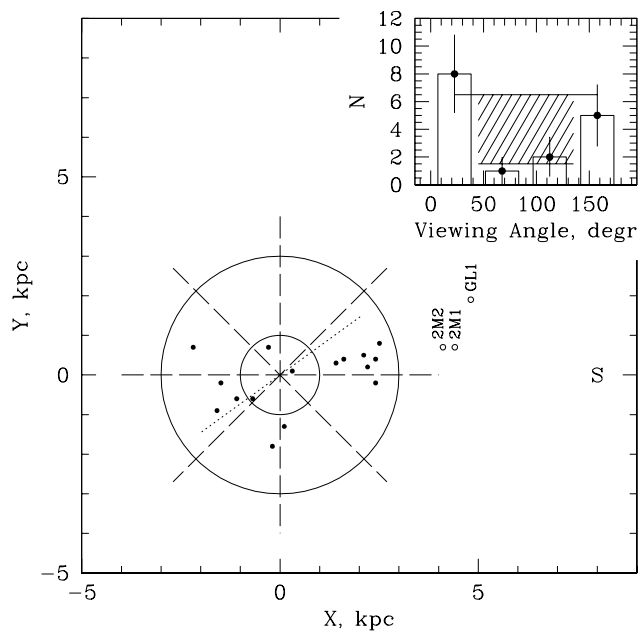}}
\caption{Spatial location of the globular clusters in 
the selected region (see 
Section~\ref{Missing_clusters_section}) projected on 
the Galactic plane. 
The Sun is marked with S, the Galactic Center lies at 
the origin of the plot. 
The X-axis increases toward the Sun, and the Y-axis 
increases in the direction of Galactic rotation. 
The locations of 2MASS\,GC01, 2MASS\,GC02, and 
GLIMPSE\,C01 are indicated (2M1, 2M2, and GL1, 
respectively).
The dotted line shows the position of the Galactic bar, 
according to Weinberg (\cite{wei92}).
Dashed lines indicate the limits of the bins for the 
histogram shown in the inset - it gives the distribution 
of the number of clusters in the selected region viewed 
from the Galactic Center. 
The viewing angle is 0 deg in the direction toward the 
Sun and it increases to 180 deg in the antisolar 
direction. Poisson uncertainties are shown with bars. 
The area of the shaded zone limited by the average level 
of the outer and the inner bins corresponds to the number 
of the clusters, needed to obtain a homogeneous 
distribution.}
\label{Missing_clusters_plot}
\end{figure}

An observer at the Galactic Center should detect an equal 
number of globulars toward the Sun, in the antisolar 
direction, and in the directions perpendicular to the line, 
connecting the Sun and the Galactic Center, i.e. the 
globular cluster distribution along any direction in the 
Galactic plane should be flat. However, it show some 
structure (Figure~\ref{Missing_clusters_plot}): there are 
more clusters in the directions toward the Sun and in the 
opposite direction than in the direction perpendicular to 
the line connected the Galactic Center and the Sun. Ten 
additional clusters are necessary to flatten the histogram 
(indicated by the shaded area), setting a lower limit of 
the missing clusters.

This estimate has to be treated with caution for a number 
of reasons. For example, it is sensitive to the adopted 
globular cluster distance, and may contain systematic 
errors related to the adopted globular cluster scale (and 
the Galactic Center distance). In addition, while the 
location of the clusters on the sky are known very well, 
their distances contain statistical observational errors, 
artificially extending the distribution along the line of 
sight, similarly to the ``finger of God'' effect known to 
the extragalactic astronomers. This would move some of the 
clusters from the middle two bins to the outer bins, 
increasing estimate. 

Finally, the spatial location of the clusters may be 
affected by the presence of a bar or a triaxial bulge in 
our Galaxy. The major axis of the bar has an angle of only 
about 36$\pm$10 deg from the line of sight toward the 
Galactic center (Weinberg \cite{wei92}), with the near end
lying in the first quadrant. The major axis of the bar is 
close to the direction in which the cluster distribution 
is elongated. 

Our result suggests that the searches for hidden clusters 
carried out so far are biased toward the regions closer
to the Galactic Center so the further away from the 
Galactic Center is a region, the less likely it is to get 
attention. This implies that the future searches has to 
modify their strategy, including regions in the Galactic 
plane, as far as 7-15 deg from the Galactic Center.

\section{Summary}

We report deep near infrared photometry of the newly 
discovered Galactic globular Cluster GLIMPSE-C01, and we 
derived for the first time the metal abundance of this 
object from the slope of the RGB. The cluster appears 
metal-poor, with [Fe/H]=$-$1.6 in the scale of Zinn. We
confirm the distance and reddening estimates of 
Kobulnicky et al. (\cite{kob05}), placing the cluster 
at D$\sim$3.7$\pm$0.8 kpc, behind A$_V$$\sim$15 mag of 
visual extinction.

Finally, we estimate the number of the missing clusters in
the central region of the Milky Way. Based on the location 
of the known clusters, and assuming radial symmetry of the 
cluster distribution around the Galactic center, we 
conclude that the Milky Way contains at least 10$\pm$3 
undiscovered globular clusters.

\begin{acknowledgements}
This publication makes use of data products from the Two 
Micron All Sky Survey, which is a joint project of the 
University of Massachusetts and the Infrared Processing 
and Analysis Center/California Institute of Technology, 
funded by the National Aeronautics and Space Administration 
and the National Science Foundation. This research has made 
use of the SIMBAD database, operated at CDS, Strasbourg, 
France. The authors thank Dr. Ortolani for the useful 
comments.
\end{acknowledgements}

\appendix

\section{Red Giant Slope -- Metallicity calibration\label{App}}

The equations given bellow superseded Eqns. 1-4 from Ivanov \& 
Borissova (\cite{iva02}) which contained an error.

\begin{equation}
(J-K_S)_0 = RGB_{ZP} + RGB_{Sl} \times M_{K_S}
\end{equation}
\begin{equation}
RGB_{Sl} = a_0^{Sl} + a_1^{Sl} \times {\rm [Fe/H]}
\end{equation}
\begin{equation}
RGB_{ZP} = a_0^{ZP} + a_1^{ZP} \times {\rm [Fe/H]}
\end{equation}
\begin{equation}
{\rm [Fe/H]} = [ (J-K_S)_0 - a_0^{Sl}\times M_{K_S} - 
      a_0^{ZP} ] ~/~ [ a_1^{Sp} \times M_{K_S} + a_1^{ZP} ]
\end{equation}

\end{document}